%% file: bug-injector.tex
\documentclass[conference]{IEEEtran}
\usepackage{etex}
\usepackage[utf8]{inputenc}
\usepackage[T1]{fontenc}
\usepackage{microtype}
\usepackage{fixltx2e}
\usepackage{graphicx}
\usepackage{epstopdf}
\usepackage{longtable}
\usepackage{multirow}
\usepackage{makecell}
\usepackage{float}
\usepackage{wrapfig}
\usepackage{caption}
\usepackage{subcaption}
\usepackage{rotating}
\usepackage[normalem]{ulem}
\usepackage{amsmath}
\usepackage{textcomp}
\usepackage{marvosym}
\usepackage{wasysym}
\usepackage{amssymb}
\usepackage{hyperref}

\usepackage{enumitem}
\usepackage{xspace}
\usepackage{xcolor, colortbl}
\usepackage{adjustbox}
\usepackage{booktabs}
\usepackage{fp}
\usepackage[group-separator={,}]{siunitx}
\usepackage{algorithm}
\usepackage{algpseudocode}
\usepackage{smartdiagram}
\usepackage{threeparttable}
\usepackage{tablefootnote}
\usepackage{url}

\algblockdefx[DO]{Do}{EndDo}[1]{\textbf{do} #1 \textbf{times}}{\textbf{end do}}
\input{texfiles/colors}
\input{texfiles/listings}
\input{texfiles/tikz}

\def\cca#1{\cellcolor{black!#1}\ifnum #1>50\color{white}\fi{#1}}
\def\hilightcell#1{\cellcolor{black!20}#1}

\newcommand\bi[0]{\textsc{Bug-Injector}\xspace}
\newcommand{\hide}[1]{}

\renewcommand{\paragraph}[1]{\smallskip\noindent\textbf{#1.}~}
\newif\ifblind
\blindfalse % remove comment for non-blind

\newif\ifacm
\acmfalse % remove comment for non-ACM

\newif\iftight
\tightfalse % remove comment for non-tight

\begin{document}

\title{Automated Customized Bug-Benchmark Generation}

\author{
  \IEEEauthorblockN{Vineeth Kashyap,
  Jason Ruchti, Lucja Kot, Emma Turetsky, Rebecca Swords, \\
   Shih An Pan, Julien Henry, David Melski, and Eric Schulte}
    \IEEEauthorblockA{
    GrammaTech, Inc., Ithaca, NY 14850 \\
   \texttt{\{vkashyap,jruchti,lkot,turetsky,rswords,span,jhenry,melski,eschulte\}@grammatech.com}
  }
}

\maketitle
\thispagestyle{plain}
\pagestyle{plain}

\begin{abstract}
We introduce \bi{}, a system that automatically creates benchmarks for customized evaluation of static analysis tools.
We share a benchmark generated using \bi{} and
illustrate its efficacy by using it to evaluate the recall of two leading open-source static analysis tools: Clang Static Analyzer and Infer.

\bi{} works by inserting bugs based on bug templates into real-world host
programs.
It runs tests on the host program to collect dynamic traces, searches the traces for a point where the state satisfies the preconditions for some bug template, then modifies the host program to ``inject'' a bug based on that template.
Injected bugs are used as test cases in a static analysis tool evaluation benchmark.
Every test case is accompanied by a program input that exercises the injected bug.
We have identified a broad range of requirements and desiderata for bug benchmarks; our approach generates on-demand test benchmarks that meet these requirements. 
It also allows us to create customized benchmarks suitable for evaluating tools for a specific use case (e.g., a given codebase and set of bug types). 

Our experimental evaluation demonstrates the suitability of our generated benchmark for evaluating static bug-detection tools and for comparing the performance of different tools.
\end{abstract}

\ifacm
\else
\fi

\ifacm
\maketitle
\else
\fi

% \fi
%\IEEEpeerreviewmaketitle

\input{texfiles/intro.tex}
\input{texfiles/related-work.tex}
\input{texfiles/bi-methodology.tex}
\input{texfiles/foundations.tex}

\input{texfiles/test-suite.tex}
\input{texfiles/evaluation.tex}
\input{texfiles/threats-validity.tex}
\input{texfiles/conclusion.tex}

\input{texfiles/acks.tex}

\bibliographystyle{IEEEtran}
\bibliography{bib/bibliography}

\end{document}

%% file: texfiles/colors.tex
\usepackage{xcolor}

\definecolor{gt@red}{RGB}{216,30,5}
\definecolor{gt@dkred}{RGB}{128,0,0}

\definecolor{gt@gray}{RGB}{110,124,146}
\definecolor{gt@dkgray}{RGB}{98,110,128}

\definecolor{gt@yellow}{RGB}{255,128,0}
\definecolor{gt@blue}{RGB}{0,68,153}

\definecolor{gt@plum}{HTML}{75507b}

\definecolor{dkpurple}{RGB}{153,0,204}
\definecolor{dkblue}{RGB}{0,0,204}
\definecolor{dkgreen}{rgb}{0,0.5,0}
\definecolor{dkred}{rgb}{0.5,0,0}
\definecolor{gray}{rgb}{0.5,0.5,0.5}

%% file: texfiles/listings.tex
\usepackage{listings}

\definecolor{dkgreen}{rgb}{0,0.5,0}
\definecolor{dkred}{rgb}{0.5,0,0}
\definecolor{gray}{rgb}{0.5,0.5,0.5}

\lstdefinelanguage{diff}{
  morecomment=[f][\color{blue}]{@@},     % group identifier
  morecomment=[f][\color{dkred}]-,       % deleted lines
  morecomment=[f][\color{dkgreen}]+,     % added lines
  morecomment=[f][\color{magenta}]{---}, % Diff header lines (after +,-)
}

\lstdefinelanguage
   [x64]{Assembler}     % add a "x64" dialect of Assembler
   [x86masm]{Assembler} % based on the "x86masm" dialect
   {morekeywords={CDQE,CQO,CMPSQ,CMPXCHG16B,JRCXZ,LODSQ,MOVSXD, %
                  POPFQ,PUSHFQ,SCASQ,STOSQ,IRETQ,RDTSCP,SWAPGS, %
                  rax,rdx,rcx,rbx,rsi,rdi,rsp,rbp, %
                  r8,r8d,r8w,r8b,r9,r9d,r9w,r9b}} % etc.

\lstdefinelanguage{JavaScript}{
  keywords={typeof, new, true, false, catch, function, return, null, catch, switch, var, if, in, while, do, else, case, break},
  keywordstyle=\color{dkred},
  ndkeywords={class, export, boolean, throw, implements, import, this},
  ndkeywordstyle=\color{dkred},
  identifierstyle=\color{black},
  sensitive=false,
  comment=[l]{//},
  morecomment=[s]{/*}{*/},
  commentstyle=\color{dkgreen},
  stringstyle=\color{gray},
  morestring=[b]',
  morestring=[b]"
}

%% file: texfiles/tikz.tex
\usepackage{tikz}

\usepackage{pgfgantt}

\usetikzlibrary{
    shapes
  , shadows
  , trees
  , arrows
  , fit
  , calc
  , positioning
  , backgrounds
  , decorations
  , decorations.pathmorphing
  , decorations.text
  , decorations.pathreplacing
}

\tikzstyle{mk} = [draw, fill=gt@red!20, rounded corners,
  minimum height=5.5em, text width=4em, text centered]
\tikzstyle{src} = [draw, fill=gt@yellow!20, rounded corners,
  minimum height=5.5em, text width=4.5em, text centered]
\tikzstyle{exe} = [draw, fill=gt@yellow, rounded corners,
  minimum height=4ex, text width=4em, text centered, text=white]
\tikzstyle{meta} = [draw, fill=gt@blue!80, rounded corners,
  minimum height=4ex, text width=4em, text centered, text=white]
\tikzstyle{var} = [exe, fill=gt@dkred, text=white]
\tikzstyle{db} = [exe, fill=gt@blue, text=white, minimum height=6ex]

\tikzstyle{stage} = [draw, thick, text=white, inner sep=1.5ex]

\tikzstyle{move} = [->,very thick, color=gt@dkgray]
\tikzstyle{connect} = [move,<->]

%% file: texfiles/intro.tex
\section{Introduction}
\label{sec:intro}

%% 'exist today' is more natural at end of sentence.
Several static analysis tools for finding bugs in programs exist today.
%% Break sentence entirely at semicolon, or do ``; however,''-> ``, but''
Researchers in academia and industry are constantly working on creating new tools and sophisticated techniques for static bug finding.  
However, evaluating static analysis tools remains a challenge. 
%A comprehensive evaluation should consider various aspects of the tools, including, the \emph{scope} (e.g., kinds of bugs detected, language features supported), \emph{performance characteristics} (e.g., precision, recall, scalability, incrementality), \emph{user experience} (e.g., warning quality and ranking, integration with developer workflow, determinism of results, configurability, setup overhead on new codebases), and \emph{surrounding facets} (e.g., tool extensibility and support, tool updates, documentation). 
A good evaluation system will guide impactful improvement in bug-finding tools by identifying their blind spots, furthering their adoption and effective use. 

%While our work can be more broadly applicable for static analysis evaluation,
%% 'aspect of evaluating'->'evaluation metric for'
%% ``the \emph{recall of a tool''->''\emph{recall}''.
There are multiple aspects of static analysis tools that are important to evaluate. In this paper, however, we mainly focus on one key evaluation metric for static analysis tools: \emph{recall}. 
Virtually all static analysis tools used widely for bug detection on C/C++ programs are unsound~\cite{livshits2015in}. 
Measuring the recall of a tool helps understand the degree to which it is unsound. 
Answering ``how well can a tool find all the bugs in a program'', i.e., recall, in a convincing manner is difficult.
It is hard---if not impossible---to enumerate all bugs in any non-trivial program. However, we can estimate the recall of a tool by counting how many \emph{previously-known bugs} in a given set of programs are found by the tool. Such estimated recall rates can be particularly useful for \emph{comparing} different tools or tool configurations.
There is a large body of previous work~\cite{juliettestsuite,stonesoup,Shiraishi2015,webgoat,wilander2002comparison,wilander2003comparison,Newsham2006,BugZoo,lu2005bugbench,nilson13bugbox,securibench,Zitser2004,nistsard,pewny2016evilcoder,dolan2016lava,Roy2018} on creating benchmarks containing known bugs.
Despite this significant progress, a recent study by Delaitre et al. ~\cite{delaitre15evaluating} found that there is still a shortage of test cases for evaluating static analysis tools and thus a need for real-world software with ground-truth information about known bugs.
% We believe that \bi{} addresses this need. 

% Unnecessary requirements(?) Maybe you want the bugs to be caught before running the tool. We want to flexible regarding. 
% Not all the requirements/goals are universal. Consider ``span the execution lifetime of a program'' and ``manifest for a very small fraction of possible inputs''.
% While these seem like reasonable requirements on a benchmark for fuzzers, they do not necessarily apply on a benchmark for static analysis tools finding a wide variety of bugs. 

%% 'in'->'for'
To address this need, we first discuss some desirable properties for a benchmark suite that contains known bugs and is targeted towards evaluating static analysis tools. 

\paragraph{Real-world-like} The benchmark's programs should be representative of real-world programs (e.g., in size and complexity).  % Analysis tools should be evaluated on the kinds of programs they are expected to be run on during software development or audit.
% The benchmark's bugs should be representative of real-world bugs (e.g., in content and integration into the host).

% \bi-generated benchmarks satisfy the {\bf realism and statistical significance} requirement in several ways. \bi takes real-wold host programs and injects real-world bugs into dynamically determined locations which result from the native control and data flow of the host programs. This creates code that is both buggy and realistic.
% Manually curated benchmarks such as the Juliet suite are typically not realistic in terms of quantity or quality of software.

\paragraph{Reliable ground truth} Each known bug in the benchmarks should manifest on at least one execution of the program. If they do not, any recall estimate based on the benchmarks is not meaningful. Ideally, each known bug should come with a proof-of-existence, such as an input that can trigger the bug.

% In addition, the manual collation of a defect corpus is error-prone, and corpora users sometimes discover that individual test cases do not contain the bugs they claim to contain \cite{juliet13release}.

% In addition, the level of guarantee provided as to the actual existence of bugs -- let alone exploitable vulnerabilities -- varies. LAVA programs come with an input to trigger the bug and are validated to check that they return exit codes associated with buffer overflows. In the LAVA benchmark \cite{lavacorpora}, each buggy program also comes with a \texttt{gdb} backtrace demonstrating the bug. On the other hand, EvilCoder programs do not come with any analogous guarantees.

% \bi{} test cases all come with a ground truth and specific inputs which can generate dynamic traces that trigger the bug. So do LAVA test cases, as we have explained already. On the other hand, EvilCoder benchmarks do not come with any guarantees of the existence of an exploitable bug, which makes it difficult to measure recall accurately. Juliet tests do come with a ground truth, although this ground truth is collated by humans and has been found to contain errors \cite{juliet13release}.

%% This has noticeably different structure to the previous points, which are all ``X should Y''
%% maybe ``The benchmark suite should be generated automatically to eliminate...''
%% or ``Benchmark suite generation should be automated ...''
\paragraph{Automated generation} To eliminate staleness, the tests in the benchmark suite should be generated automatically: on demand, without manual effort, in the quantity desired for statistical significance.

% Hand-curated defect corpora \cite{nistsard,stonesoup,lu2005bugbench,nilson13bugbox,webgoat,securibench} are expensive and time-consuming to assemble. Consequently, they are typically relatively small and infrequently updated. 
% An automated approach overcomes the problem of the small size and potential staleness of test suites as new tests can be generated on demand in the quantity desired.

% To remedy the above situation, there has been interest in generating bug-containing suites automatically with systems such as EvilCoder \cite{pewny2016evilcoder} and LAVA \cite{dolan2016lava,lavacorpora}.  However, automated approaches potentially suffer from their own drawbacks. 

% Test cases should be non-static and easy to generate on-demand  \cite{dolan2016lava, pewny2016evilcoder}
% \bi{} can generate as many test cases as desired modulo the availability of suitable host programs and bug templates. 

%% ditto: ``Benchmarks should be customizable to users' differing analysis needs and codebases...
%%  or ``Users should be able to {customize benchmarks,generate customized benchmarks} that are tailored to their codebases and analysis needs''
%% (or some combination)
\paragraph{Customizable}
Users should be able to generate customized benchmarks that are tailored to their codebases and bug distribution expectations: one fixed benchmark does not suit all. For example, Herter et al.~\cite{Herter2017} suggest that certain sectors (such as the aerospace and automotive industries) deem recursive function calls inappropriate. %% Clearly, the same restriction does not apply to all industries: some organizations will have codebases that \emph{do} include recursion, and they will want static analysis tools to reason about recursive functions. 
%% ``Also''->''Similarly''
Similarly, not all bug classes are equally important to all users of static analysis tools.
%% Does this sentence say anything new?
%Thus, it is useful to create a customized benchmark targeting the user's codebase and bug distribution expectations. % potentially discuss What Developers Want and Need from Program Analysis: An Empirical Study, that different users want to use analysis tools finding different kinds of bugs

%% ``bugs of a wide variety of bug types'' -> ``a {wide,broad} variety of bugs''.
%% in this context ``disregarding certain types of bugs'' appears to be contradicting the main point of the paragraph. Maybe recast as ``Users can [choose to] customize their evaluation by specifying the types of bugs to {consider,insert,...}.''
\paragraph{Broad coverage of bug types} Benchmarks should include a broad variety of bugs: for example, bugs corresponding to a large set of different Common Weakness Enumeration entries (CWEs)~\cite{cwe}. Users can choose to customize their evaluation by disregarding certain types of bugs.

% The techniques used to create test cases may limit the benchmarks to specific kinds of weaknesses: buffer overruns in the case of LAVA and taint-related weaknesses for EvilCoder. 

% \bi is applicable for many CWE classes \cite{cwe} and allows injection into many kinds of software.

% However, synthetic benchmarks do cover multiple CWEs.

% In addition, both LAVA and EvilCoder tools are limited in the type of bugs they insert: LAVA works only with buffer overflows, EvilCoder with taint-based weaknesses. 

%% is the benchmark supposed to do the comparison/contrast automatically or just provide information to support a human doing those things? This says/implies the former but I think you might mean the latter
\paragraph{Suitable for usefully evaluating and comparing the recall of static analysis tools} Comparing the recall on the benchmarks should discriminate between static analysis tools. The benchmark tests should provide guidance to further improve the recall of a given tool, e.g., by including bugs which are within scope for the tool, but which the tool is unable to detect.

%% This is the only ``must'' in a list that is otherwise ``should'' - perhaps it should be at the top.
\paragraph{Implemented independently of evaluated techniques} To avoid circularity, the techniques used to create the benchmark suite should be independent of the techniques the benchmark suite will be used to evaluate.  Otherwise evaluations using the benchmark suite will be biased by hiding shared limitations.

We address all of the above desired properties through \bi{}, a system that automatically generates benchmarks containing known bugs. \bi{}-generated benchmarks have a broad range of applications, but the one we present in this paper is particularly suited to estimating and comparing the recall rates of static analysis tools, such as the open-source tools Clang Static Analyzer and Infer.

\bi{} starts from (i) a set of \emph{bug templates} (\autoref{bug-templates}) that represent known bugs, (ii) a \emph{host program}, i.e., an existing real-world software application, and (iii) a set of tests to exercise the host program.
It searches dynamic traces of the host program to identify \emph{injection points} where the state satisfies a bug template's preconditions.
Using dynamic state to identify bug injection locations, rather than using information from static analyses, provides independence from bias and from the limitations of static analysis techniques (such as pointer analysis imprecision or SMT solver weaknesses).
For each of the identified injection points (or a random subset thereof), \bi{} creates a new variant of the host program by inserting a bug based on the bug template, integrating with existing data and control flow.
\bi{} relies on existing data and control flow complexity in the host programs to generate realistic contexts for injected bugs.
\bi{} outputs multiple versions of each host program, each containing one injected bug and identifying a concrete program input that will trigger the injected bug.

%% What is the intent of this sentence? If you mean that bi is independent (or largely independent, or something) of these factors, say that explicitly.
%% 'a variety' might be a bit weak. maybe
%% ``a variety of bug classes'' -> ``a broad range of bug classes''
%% and
%% ``a variety of programs''->''programs of various sizes, functionality types, and complexity levels''
\bi{} can inject bugs from a broad range of bug classes into programs of various sizes, functionality types, and complexity levels (\autoref{sec:testsuite}). This customizability allows for additional uses of \bi{} beyond tool evaluation. For instance, a tool developer who creates analysis checkers for a new kind of bug can use \bi{} to generate test cases containing bugs of that kind to quickly evaluate the checker's recall against real-world software (\autoref{subsec:example-injection}). 
This usage of \bi{} can complement the typical method of testing analysis checkers with small, hand-crafted tests.

The specific contributions of this paper are:
\begin{enumerate}[noitemsep,nolistsep]
  \item The \bi{} system, a novel technique to automatically generate customized, realistic benchmarks with known bugs that can be triggered using accompanying inputs. We describe \bi{}'s architecture, functionality, and underlying algorithms in \autoref{sec:bi-arch}. As far as we know, \bi{} is the only existing system (\autoref{tab:related-work}) capable of producing benchmarks that meets all desired properties discussed above.
    %% ``we created'' --> ``we have created''
    %% ``belonging to'' -> ``corresponding to''
    %% ``CWEs'' -> ``CWE entries''
  \item Openly-available benchmark suites (\autoref{sec:testsuite}) generated using \bi{}. We have created bug templates (both manually and automatically) from different sources corresponding to a wide variety of CWE~\cite{cwe} entries and injected them into open-source real-world programs.
    %% ``can contrast''->''can be used to [compare and] contrast''
    %% ``contrasting and comparing'' feels a little unnatural to me, ymmv
  \item An extensive evaluation of two leading open-source static analysis tools for C/C++ programs---Clang Static Analyzer (CSA)~\cite{csa} and Infer~\cite{fbinfer}---on our generated benchmarks. Our results (\autoref{sec:evaluation}), show that: (a) both of these tools fail to detect bugs that are seemingly in scope for them, (b) our benchmarks can be used to compare the recall of the two tools, and (c) our benchmarks can contrast two analysis configurations of CSA, showing that \bi{} can be used to automatically tune analysis configurations customized to a codebase. We also filed bug reports for CSA and Infer on certain missed warnings. Additionally, we show that a closely related work, LAVA~\cite{lavacorpora}, is not suitable for comparing static analysis tools. 
\end{enumerate}

%% ``recall of tools'' -> ``tool recall''
In the remainder of the paper, we compare to related work (\autoref{sec:relwork}), describe challenges in estimating tool recall (\autoref{sec:foundations}),  discuss limitations and future work (\autoref{sec:threats}), and conclude (\autoref{sec:conclusion}).

%% file: texfiles/related-work.tex
\section{Related work}
\label{sec:relwork}

Creating bug-containing benchmarks for testing and evaluating bug-finding tools has attracted significant research attention in recent years. In this section, we compare \bi{} to the closest related work, summarized in \autoref{tab:related-work}.

\begin{table}[h]
  \caption{\label{tab:related-work}Summary comparing \bi{} (\textbf{BI}) with other closely related work across the different properties outlined in \ref{sec:intro}. Columns: \textbf{EC}=EvilCoder, \textbf{Synth}=Synthetic benchmarks, Wild=wild caught bugs.  Values: Ltd.=Limited, Yes*=subject to some errors.}
  \begin{tabular}{llllll}
    \toprule
    \textbf{Property}       & \textbf{BI} & \textbf{LAVA} & \textbf{EC} & \textbf{Synth} & \textbf{Wild} \\
    \midrule
    Real-world-like         & Yes         & Yes           & Yes         & No             & Yes           \\
    Reliable ground truth   & Yes         & Yes           & No          & Yes*           & Ltd.          \\
    Automated, not fixed    & Yes         & Yes           & Yes         & No             & No            \\
    Customizable            & Yes         & Yes           & Yes         & No             & No            \\
    Wide coverage of CWEs   & Yes         & No            & No          & Yes            & No            \\
    Evaluate static tools?  & Yes         & No            & No          & Ltd.           & Ltd.          \\ 
    Independent?             & Yes         & Ltd.          & No          & Yes            & Yes           \\
    \bottomrule
  \end{tabular}
  \iftight
  \vspace*{-2em}
  \fi
\end{table}

\paragraph{Synthetic benchmarks} Several efforts have targeted manual creation of  artificial test programs containing bugs. Some prominent examples are:  Juliet tests~\cite{juliettestsuite}, the IARPA STONESOUP snippets \cite{stonesoup}, Toyota ITC benchmarks~\cite{Shiraishi2015}, OWASP WebGoat~\cite{webgoat}, Wilander et al.,~\cite{wilander2002comparison,wilander2003comparison}, and ABM~\cite{Newsham2006}. However, synthetic benchmarks have limited applicability in identifying how tools perform on real-world code.

\paragraph{Wild} Bugs may be mined and curated from real-world software. Some prominent examples of such curated bug collections are: BugZoo~\cite{BugZoo}, BugSwarm~\cite{Tomassi2019}, ManyBugs~\cite{LeGoues15tse}, Defects4J~\cite{defects4j,Habib2018}, BugBench~\cite{lu2005bugbench}, BugBox~\cite{nilson13bugbox},  SecuriBench~\cite{securibench}, and Zitser et al.,~\cite{Zitser2004}. While they have the advantage of being real-world-like, they have varying degrees of ground truth, and not all of them come with proof-of-existence. There is also very little benchmark-user customizability with respect to bug type coverage and distribution. % Such benchmarks would also bias the evaluations: we are testing how well tools can find previously found bugs in the wild.

The curation of both wild and synthetic benchmarks requires substantial manual effort and is prone to errors (e.g., both the Juliet test cases and the Toyota ITC benchmarks have required corrections~\cite{juliettestsuite,Herter2017}). They are fixed and not customizable, with pre-determined target code constructs and bug types. They therefore have limited applicability for evaluating and comparing the recall of static analysis tools.
% These corpora are all relatively small and static. They provide a varying level of structure with respect to ground truth; for example, the Juliet test cases supply both vulnerable and non-vulnerable variants of code, while some of the other benchmarks do not.
SARD~\cite{nistsard} is perhaps the largest openly available collection of known buggy test programs, put together by the SAMATE group at NIST. It contains both synthetic and wild benchmarks. 

\paragraph{EvilCoder} This system~\cite{pewny2016evilcoder} uses static analysis to find sensitive sinks in a host program and connects them to a user-controlled source to inject taint-based bugs. A significant disadvantage is that there is no guarantee that inserted bugs are true positives---which makes it unsuitable for estimating recall. Indeed, the paper does not evaluate bug-finding tools on EvilCoder test cases. EvilCoder injected bugs inherit the limitations of the static analysis tools used as a part of the injection pipeline, and therefore may bias evaluation of other static analysis tools. EvilCoder is limited to taint-based bugs.

\paragraph{LAVA} This system~\cite{dolan2016lava, lavacorpora} inserts bugs into host programs by identifying situations where user-controlled input can trigger an out-of-bounds read or write. LAVA bugs come with an input to trigger the bug and are validated to check that they return exit codes associated with buffer overflows. However, this approach is limited to inserting buffer overruns, and other kinds of bugs are left as future work. More recently~\cite{Hulin2017}, LAVA has been extended to a small number of additional bug types. LAVA test cases are generated to satisfy an additional goal: the bugs must manifest only on a small fraction of all possible inputs. This requirement seems targeted towards testing fuzzing tools; we do not think it is necessarily applicable in the context of testing static analysis tools.\footnote{Many famous bugs, e.g. HeartBleed~\cite{HeartBleed}, execute on the majority of possible inputs.} To satisfy the requirement, LAVA injects bugs with certain patterns (the ``knob and trigger'' pattern which relies on magic values). It is unclear how realistic this bug pattern is with respect to bugs found in production software. A more detailed discussion of the suitability of LAVA benchmarks for static analysis evaluation is provided in~\autoref{subsec:lava}. Another closely related technique is Apocalypse~\cite{Roy2018}, which is similarly targeted towards creating challenging benchmarks for fuzzing and concolic execution tools. 

As opposed to the synthetic and wild benchmarks, EvilCoder, LAVA, and \bi{} are automated and can create large number of bugs in custom real-world programs. 

\bi{} uses bug templates and a host program to produce a suite of programs containing one known bug apiece, along with an input that can trigger each bug. The available bug templates cover a large number of CWEs, and new bug templates are easy to create. Through empirical evaluation (\autoref{sec:evaluation}), we show that \bi{} generated benchmarks are suitable for evaluating and comparing the estimated recall of static analysis tools.

Other related techniques are mutation testing~\cite{Demillo1978, Hamlet1977} and fault injection~\cite{Christakis2017,Marinescu2009,Martins2002}. 
Mutation testing is used to evaluate the quality of a test suite, and is different from our work because the mutations are much simpler, are not dynamically targeted, and are not guaranteed to introduce real bugs.
Compared to our work, fault injection techniques serve a different purpose: they aim to evaluate the robustness of software in the presence of various kinds of faults, e.g., data corruption, errors returned by library functions. 
Typically, these techniques inject or emulate faults in software at runtime, and then compare the dynamic behavior of software during normal and fault-induced runs. 
Faults injected by these techniques are fairly simple~\cite{Christakis2017}, and in contrast to \bi{}, faults are not integrated with the host program. 

%% file: texfiles/bi-methodology.tex
\section{Bug Injector}
\label{sec:bi-arch}

In this section, we describe the tooling used for \bi{}, introduce bug templates, and describe how \bi{} works. We illustrate the injection of a bug template into a host program, and discuss potential applications. 

\subsection{Tooling}
\label{supporting-tooling}

\bi{} is implemented using the Software Evolution Library
(SEL)~\cite{sel2018manual}, an open-source toolchain that provides a uniform interface for instrumenting, tracing, and modifying software.
SEL supports multiple programming languages. 
Currently, \bi{} works on C/C++, Java, and JavaScript\footnote{Java and JavaScript support is experimental, under heavy development.}  software. In this paper, we focus on \bi{} as applied to C/C++ software. 
C/C++ software modifications are implemented via Clang's libtooling API. 
Clang's libtooling provides a solid foundation for parsing and program modification in the presence of the latest C/C++ syntactic features, making \bi{} applicable to a wide range of C/C++ software.

\subsection{Bug templates}
\label{bug-templates}

\bi{} is able to inject a wide range of bug types, based on the provided bug templates. 
A bug template is defined in Common Lisp, and  it specifies: (a) the dynamic and static requirements for a successful bug injection, (b) the code snippets constituting the bug itself, and (c) how these code snippets should be integrated into the program. 
A bug template consists of one or more \emph{patches}.\footnote{A successful bug injection applies all the patches in a bug template.}   
An example bug template consisting of a single patch is provided in~\autoref{scion}.
Each patch has the following fields.

\begin{figure}[h]
  \begin{subfigure}{\columnwidth}
    \centering
\begin{lstlisting}[language=C]
void f1(char *src) {
 char *dst = 0; // 'dst' initialized to a null ptr 
 memcpy(dst + 0, src, 10); // expected warning: null 
 // ptr argument in call to memory copy function
}
\end{lstlisting}
    \caption{\label{regression} Clang Static Analyzer (CSA) regression test, annotated with expected tool behavior. This test program contains a bug: the first argument to the memory copy function is null.}
  \end{subfigure}%
  \vspace{0.3cm}
  \begin{subfigure}{\columnwidth}
    \centering
    \resizebox{0.8\columnwidth}{!}{%
    \begin{tabular}{l l}
      \toprule
      \textbf{code} & \texttt{memcpy(\$dst + 0, \$src, \$num);} \\
      \midrule
      \multirow{2}{*}{\textbf{free-variables}} & \texttt{\$dst}: pointer to char \\
                                                & \texttt{\$src}: pointer to char \\
                                                & \texttt{\$num}: integer \\
      \midrule
      \textbf{precondition} & $\mathsf{value}$\texttt{(\$dst, p) $=$ 0} $\land$ $\mathsf{value}$\texttt{(\$num, p) $>$ 1} \\ 
    % \midrule
      % \textbf{include} & \texttt{<string.h>} \\
      % \midrule
      % \textbf{postcondition} & true \\
      \bottomrule
      \end{tabular}
    }
    \caption{\label{scion} A bug template containing a single patch corresponding to the bug in \autoref{regression}. 
    For presentation, the patch has been abstracted from its Common Lisp definition.}
  \end{subfigure}
  \vspace{0.3cm}
  \begin{subfigure}{\columnwidth}
    \centering
    \begin{lstlisting}[language=diff]
/* global variable declarations */
static char *lastout;
static char *prog;
static int out_byte;
/* ... lots of code, some interact with globals */
static int grep(int fd) {
  /* ... more code */
+ /* from input (./harness.sh test BIN 1) */
+ /* POTENTIAL FLAW */
+ memcpy(lastout + 0, prog, out_byte);
  reset(fd);
  lastout=0;
\end{lstlisting}
    \caption{\label{diff}The diff resulting from injecting the bug template in~\ref{scion} into grep. Here, \texttt{\$dst} $\rightarrow$ \texttt{lastout}, \texttt{\$src} $\rightarrow$ \texttt{prog}, \texttt{\$num} $\rightarrow$ \texttt{out\_byte}.}
  \iftight
  \vspace*{-1em}
  \fi
  \end{subfigure}%
  \caption{\label{fig:scion-injection}A CSA regression test (\ref{regression}), a corresponding bug template (\ref{scion}) created manually, and the diff resulting from injection of
  this bug template into the grep program (\ref{diff}).}
  \iftight
  \vspace*{-1em}
  \fi
\end{figure}
 
\paragraph{\texttt{code}} The buggy code that will be inserted into the host program. In~\autoref{scion}, the buggy code is a call to \texttt{memcpy}. 
The buggy code can contain references to free variables.

\paragraph{\texttt{free-variables}} A list of type-qualified free variables in the buggy code. These are matched to type-compatible in-scope variables at the injection location in the host program. 
Occurrences of the free variables in \textbf{\texttt{code}} are replaced with the matched host program variables  before injection. In \autoref{scion}, the free variable listing implies that {\tt \$dst} and {\tt \$src} should be bound to host program variables with type \texttt{char*}. 

\paragraph{\texttt{precondition}} Any boolean predicate constructed using the following primitives defined over the in-scope variables (\texttt{\$v}) at a program point (\texttt{p}) in the dynamic trace: 
\begin{itemize}
  \item $\mathsf{value}$\texttt{(\$v, p)}: the value of \texttt{\$v} at \texttt{p}
  \item $\mathsf{size}$\texttt{(\$v, p)}: the dynamically allocated size of memory pointed to by \texttt{\$v} at \texttt{p} 
  \item $\mathsf{ast}$\texttt{(p)}: the abstract syntax tree at \texttt{p} 
  \item $\mathsf{name}$\texttt{(\$v)}: the name of \texttt{\$v}
  \item $\mathsf{type}$\texttt{(\$v)}: the static type of \texttt{\$v}
\end{itemize}
The primitives $\mathsf{value}$ and $\mathsf{size}$ allow matching on dynamic conditions, whereas $\mathsf{ast}$, $\mathsf{name}$, and $\mathsf{type}$ allow matching on static conditions.  
\bi{} uses the precondition predicate to search the dynamic traces for suitable injection locations: points in the trace that meet the precondition. 
The input that gives rise to a trace is called the ``witness'' of that trace. 
The buggy code injected into the source at the precondition-matching location will be executed when run with the witness input. 
In \autoref{scion}, the precondition specifies that at an injection point \texttt{p}, (a) an in-scope variable bound to \texttt{\$dst} is a null pointer, and (b) another in-scope variable bound to \texttt{\$num} has a value $> 1$.

% \paragraph{\texttt{postcondition}} An optional boolean predicate that must hold after the buggy code has been exercised. This predicate is used to validate dynamically that the witness triggers the bug.  
%A postcondition ``true'' always holds as long as the preconditions have been met. 
% If no postcondition is specified (or equivalently, a trivial ``true'' postcondition is specified, as in the example in \autoref{scion}), no additional validation is performed. 
% Else, validation instrumentation is inserted to validate that the postcondition holds. 
% All validation instrumentation is removed before delivery of the buggy program.

The example bug template in~\autoref{scion} was manually created based on an existing regression test (\autoref{regression}) for CSA. 
This regression test contains a bug at the call to \texttt{memcpy}: that its first argument is a null pointer. 
A successful injection of the bug template in~\autoref{scion} will insert a call to \texttt{memcpy}, where (a) \texttt{\$src}, \texttt{\$dst}, and \texttt{\$num} are replaced with host program variables, and (b) the variable bound to \texttt{\$dst} is null and the variable bound to \texttt{\$num} is $>1$, before the call to \texttt{memcpy}. 
Thus, the bug injection attempts to create the same kind of bug, but embedded and integrated with the host program's data and control flow complexity. 
The null value of the host program variable bound to \texttt{\$dst} comes from an existing sequence of host program events (i.e., it is not artificially generated): e.g., the pointer might have been set to null at a distant program point in a different function, or its value copied from some other pointer which happens to be null under certain conditions triggered by the witness input.  

Typically, creating a bug template from an example bug requires identifying (a) the relevant buggy code, (b) the free variables in the buggy code that must be re-bound in the host program, and (c) preconditions to ensure the bug is successfully transferred to the host program.

% To illustrate this new workflow, we take an existing regression test (\autoref{regression}) for a Clang Static Analysis (CSA) checker. The test program includes comments regarding where and what kind of warnings are expected when the CSA tool is run on the test program---as expected, CSA reports a warning at the appropriate location. 
% To see how well CSA finds the same kind of bug in a real-world program, we convert the regression test into a bug template containing a single patch, as given in \autoref{scion}. 

% The precondition requires that at the host program injection point, the host program variable mapped to the free variable \texttt{dst} is a null pointer---thus the same kind of bug (null pointer argument in call to memory copy function) is exhibited in the host program after injection. 

\subsection{Technique}

\begin{figure}[h]
  \vspace*{-1ex}
  \begin{center}
    \adjustbox{max width=0.5\textwidth}{\input{texfiles/pipeline}}
  \end{center}
  \vspace*{-2ex}
  \caption[\bi overview]{\label{overview}\bi pipeline.}
  \iftight
  \vspace*{-1em}
  \fi
\end{figure}

The \bi{} pipeline of {\em instrument}, {\em execute},
and {\em inject} is shown in \autoref{overview} and
described in the algorithm in \autoref{bi-algorithm}. \bi takes three inputs:
(1) a host program, (2) a set of tests for this program, and
(3) a set of bug templates. It attempts to inject bugs from the
set of bug templates into the host program, and returns multiple different buggy versions of the host program.  Each returned buggy
program variant has at least one known bug (the one that was injected), and is associated with a \emph{witness}---a test input which is known to exercise the injected bug. 

\begin{figure}
  \input{texfiles/bi-algorithm}
  \caption{\label{bi-algorithm}\bi algorithm.}
  \iftight
  \vspace*{-1em}
  \fi
\end{figure}

The \bi{} algorithm begins by instrumenting the host program (\autoref{bi-algorithm}, line~\ref{alg:inst}). The
$\mathsf{Instrument}$ method rewrites the source code of the host
program, inserting code to emit dynamic trace output. 
Traces include the values of all in-scope variables (currently limited to primitive types and pointers) at every program statement.
% The trace format can store program locations, variables, and auxiliary data. 
The algorithm then runs the instrumented program with test inputs.
The collected traces (capped by size \textit{MaxTrace}) and the test inputs that produced them are stored efficiently in a persistent binary-format database, $\mathit{TraceDB}$ (line~\ref{alg:trace}).
%\TODO{anonymize}:  Implementation details for instrumentation and tracing can be found in the SEL manual \cite{seltracingdocs}.

\bi{} then attempts to inject each of the bug templates \textit{NumInjection} times. 
For every bug template, it uses $\mathsf{Match}$ to search $\mathit{TraceDB}$ for candidate program point sets that correspondingly match the preconditions for all the patches in that bug template. 
$\mathsf{Match}$ returns a list of candidates: each candidate is a tuple of points---one program point per patch in the bug template---and a witness input. 
\bi{} randomly samples \textit{NumInjection} candidates for injection.
The candidates picked for injection are then used by $\mathsf{Inject}$ (line~\ref{alg:inject}), which takes the code in the patches of the bug template and rewrites the source code locations associated with each of the \textit{Points}. 
Source rewriting involves inserting the associated code snippet into the host program, then renaming all the free variable names with the precondition-matching and type-compatible in-scope variables of the host program.

To validate an injection, \bi{} adds instrumentation\footnote{The validation code is removed before adding it to the benchmark.} to the modified program to dynamically check that the pre-conditions hold before the injected bug upon re-execution against
$\mathit{Witness}$ (line~\ref{alg:validate}).  
The buggy program $\mathit{Bugged}$ and its associated $\mathit{Witness}$ are added to the output $\mathit{Bench}$ (line~\ref{alg:valid-add}).  
%
% This process continues until the requested number of bugs have been injected, the algorithm then returns the current $\mathit{Bench}$.
% For simplicity of presentation, the algorithm given here assumes that the requested number of bugs is far fewer than the number of dynamic trace points that satisfy the bug template preconditions.
After exhausting the given number of injections, or when no more candidate injection points are available, $\mathit{Bench}$ is returned.

As an example, consider the injection of bug template given in \autoref{scion} into the C program \texttt{grep}, resulting in the buggy variant of \texttt{grep} shown in \autoref{diff}. 
In this instance, \bi{} uses host program's global static variables, \texttt{lastout}, \texttt{prog}, and \texttt{out\_byte}, in the call to \texttt{memcpy}. For diagnostic purposes, the injected buggy code is optionally preceded by a comment that includes the input witness for the injected bug. 
When buggy \texttt{grep} is run with this input, the value of \texttt{lastout} is null before the call to \texttt{memcpy}, at least once during program execution.
This injection successfully violates the ``\texttt{memcpy} should not be called with its first argument being a null pointer'' rule, but in a different code context. 

CSA emits a warning about the bug in their regression test~\autoref{regression}.  
However, CSA fails to report a warning for the similar injected bug in this buggy version of \texttt{grep}. 
CSA has ``lost'' this bug due to its injection into a more complex context.

\subsection{Uses of \bi{}}
\label{subsec:example-injection}

% \begin{figure}
%   \centering
%   \begin{subfigure}{.5\columnwidth}
%     \centering
%     \scalebox{0.6}{
%       \smartdiagramset{planet color=orange!10, 
%       % bubble center node size=0.25cm, 
%       distance planet-satellite=2.8cm
%       }
%       \smartdiagram[connected constellation diagram]
%       {Develop Checker,Craft Test Programs,Deploy to\\ Production,Identify Problems}
%       } 
%     % \caption{\label{fig:checker-dev-current}Current static analysis checker development workflow.}
%   \end{subfigure}%
%   \begin{subfigure}{.5\columnwidth}
%     \centering
%     \scalebox{0.6}{
%     \smartdiagramset{planet color=orange!10, 
%     % bubble center node size=0.25cm, 
%     distance planet-satellite=2.8cm,
%     set color list={red!80, cyan!80, blue!30}
%     }
%     \smartdiagram[connected constellation diagram]
%     {Develop Checker,Craft Bug Templates,Test on Benchmarks,Identify Problems}
%     }
%     % \caption{\label{fig:checker-dev-new}Proposed static analysis checker development workflow using \bi{}.}
%   \end{subfigure}
%   \caption{\label{fig:checker-dev}Contrasting static analysis checker development workflows: the current workflow is presented on the left, and the proposed workflow using \bi{} is on the right.}
% \iftight
% \vspace*{-1em}
% \fi
% \end{figure}

One of the applications of \bi{} is to provide feedback to static analysis tool developers regarding the false-negative rates of their ``checkers'' on real-world programs. 
A typical workflow for building static analysis checkers\footnote{This workflow is informed by the author's discussions with static analysis tool developers and by the static analysis checker development tutorial for Phasar~\cite{PhasarTutorial}.} is an iterative process:
(1) develop a checker to detect violations of a program property, (2) test the static analysis checker on some manually crafted test programs, (3) deploy the checker into production, (4) identify failures and false-negative corner cases for the checker, (5) iterate and improve the checker. 
\bi{} can improve and accelerate this process.  Instead of manually crafting test cases, we can craft relevant bug templates. 
\bi{} can then generate checker benchmarks by injecting these bug templates into real-world programs. 
The static analysis checker can then be tested on the generated benchmarks to obtain early feedback regarding the checker's performance (such as expected false-negative rate, scalability), before deploying the checker into production.

Another application of \bi{} is customized evaluation of static analysis tools, as we have done in~\autoref{sec:evaluation}. 
We also provided the SAMATE group at NIST with \bi{}.
This group is conducting SATE VI~\cite{nistsate}: the sixth iteration of Static Analysis Tool Exposition. 
SATE is a non-competitive study of static analysis tool effectiveness, aiming at improving tools and increasing public awareness and adoption. SATE VI is already making use of \bi{} generated test programs, in addition to manually crafted test programs. 
Further, NIST is expecting to make extensive use of \bi{} for SATE VII, the next iteration of SATE. To quote the initial experience of the NIST team with \bi{}: ``using \bi{} to generate benchmarks is much faster (at least five times as
fast) than using our current manual benchmark generation process.''
For SATE VI, the participating static analysis vendors can compare how well they perform on \bi{} generated benchmarks vs. the manually created benchmarks, which will be a useful broader study regarding the effectiveness of \bi{}. 
NIST also plans to add \bi{} generated tests to the SARD dataset~\cite{nistsard}.

%% file: texfiles/pipeline.tex
\tikzstyle{move} = [->,very thick, color=gt@dkgray]%
\tikzstyle{src} = [draw, fill=gt@yellow!30, rounded corners,%
  minimum height=4em, text width=4em, text centered]%
\tikzstyle{inputs} = [src, fill=gt@blue!30, minimum height=3em, text width=7em]%
\tikzstyle{vuln} = [src, fill=gt@blue!30, minimum height=3em, text width=7em]%
\tikzstyle{action} = [font=\small\it]%
\begin{tikzpicture}[node distance=8ex]
  \node[src, fill=gt@blue!30] (host) {Host Program};

  \node[right=2ex of host] (inst) {Instrument};

  \node[text centered, right=2ex of inst] (trace) {Execute};

  \node[text centered, right=4ex of trace] (inject) {Inject \& Validate};

%% Buggy programs  
  \node[src, minimum height=5em, fill=gt@red!30, right=2ex of inject] (output) {};
  \node[src, minimum height=5em, fill=gt@red!30, below right=0.5ex of output.north west] (output2) {};
  \node[src, minimum height=5em, fill=gt@red!30, below right=0.5ex of output2.north west] (output3) {};
  \node[src, minimum height=5em, fill=gt@red!30, below right=0.5ex of output3.north west] (output4) {Buggy Programs + Witnesses};

%  \node[src, fill=gt@red!30, right=4ex of output] (witness) {Witness\\ Input};

  %% Bug Templates
  \node[vuln, below right=2ex and -3ex of inject.south west] (vuln1) {};
  \node[vuln, below right=0.5ex of vuln1.north west] (vuln2) {};
  \node[vuln, below right=0.5ex of vuln2.north west] (vuln3) {};
  \node[vuln, below right=0.5ex of vuln3.north west] (vuln4) {Bug Templates};

  %% Inputs
  \node[inputs, below left=2ex and 3ex of trace.south east] (input1) {};
  \node[inputs, below right=0.5ex of input1.north west] (input2) {};
  \node[inputs, below right=0.5ex of input2.north west] (input3) {};
  \node[inputs, below right=0.5ex of input3.north west] (input4) {Program Inputs};

  %% Draw
  \draw[move,->] (host) to (inst);
  \draw[move,->] (host) to [out=20,in=160] (inject);
  \draw[move,->] (inst) to (trace);
  \draw[move,->] (trace) to
    node[auto,above,action] {Trace}
    node[auto,below,action] {Points}
    (inject);
  \draw[move,->] (inject) to (output);
  \draw[move,->] (input3) to (trace);
  \draw[move,->] (vuln3) to (inject);
%  \draw[move,->] (inject) to (witness);

\end{tikzpicture}

%% file: texfiles/bi-algorithm.tex
\begin{algorithmic}[1]
\small
\item[{\textbf{Input: }} {Host Program, $\mathit{Host}$: $\mathit{Program}$}]
\item[{\textbf{Input: }} {Program Inputs, $\mathit{Suite}$: $\{\mathit{Test}\}$}]
\item[{\textbf{Input: }} {Bug Templates, $\mathit{Templates}$: $\{\mathit{BugTemplate}\}$}]
  \item[{\textbf{Parameters: }} {$\mathit{NumInjections}$, $\mathit{MaxTrace}$}]
  \item[{\textbf{Output: }} {Buggy Program Versions, $\mathit{Bench}$: $\{\mathit{\langle Program, Test \rangle}\}$}]

  \State {\bf let} $\mathit{Bench} \leftarrow \emptyset{},\, \mathit{TraceDB} \leftarrow \emptyset{}$

  %% Instrumentation
  \State {\bf let} $\mathit{Inst} \leftarrow \mathsf{Instrument}(\mathit{Host})$\label{alg:inst}\Comment{Instrument}

  %% Evaluation
  \For{$\mathit{Input} \in \mathit{Suite}$}\Comment{Execute}
    \State $\mathit{TraceDB} \leftarrow \mathit{TraceDB} \cup \mathsf{Collect}(\mathit{Inst}, \mathit{Input}, \mathit{MaxTrace})$\label{alg:trace}
  \EndFor

  %% Injection
%  \State {\bf let} $\mathit{Bug},\, \mathit{Point},\, \mathit{Witness},\, \mathit{Bugged}$

  \For{$\mathit{Template} \in \mathit{Templates}$}\label{alg:template-loop}
  \State $\mathit{Candidates} \leftarrow \mathsf{Match}(\mathit{Template.precondition},\, \mathit{TraceDB})$
  \State $\mathit{Sampled} \leftarrow \mathsf{RandomSample}(\mathit{Candidates}, \mathit{NumInjections})$
    \For{$\langle \mathit{Points}, \mathit{Witness} \rangle \in \mathit{Sampled}$}\label{alg:injection-loop}
    \State $\mathit{Bugged} \leftarrow \mathsf{Inject}(\mathit{Host},\, \mathit{Points},\, \mathit{Template})$\label{alg:inject}\Comment{Inject}      
    \If { $\mathsf{Validate}(\mathit{Bugged},\, \mathit{Witness})$ }\label{alg:validate}\Comment{Validate}
      \State $\mathit{Bench} \leftarrow \mathit{Bench} \cup \{(\mathit{Bugged},\, \mathit{Witness})\}$\label{alg:valid-add}
    \EndIf
    \EndFor
  \EndFor
  \State \Return{ $\mathit{Bench}$ }
\end{algorithmic}

%% file: texfiles/foundations.tex
\section{Estimating static analysis recall}
\label{sec:foundations}

As previously discussed in \autoref{sec:intro}, it is difficult to compute the exact recall of a tool. Thus, \bi{} (as well as all other related work) estimates the recall of a static analysis tool using the set of known bugs in a given benchmark, which is a subset (possibly strict) of all the bugs actually present in that benchmark. The set of known bugs in a given benchmark is referred to as the \emph{ground truth} for the benchmark. In this section, we discuss some practical issues in representing ground truth for the purposes of evaluating static analysis tools. % These are not unique to \bi{}, and have been previously mentioned.

\paragraph{Ground truth accuracy} That is, each bug in the provided list must manifest in at least one execution of the program. LAVA~\cite{dolan2016lava} provides backtraces for each test case showing that the bugs included are real. EvilCoder \cite{pewny2016evilcoder}, however, provides no such guarantees. \bi{} benchmarks come with inputs which can generate dynamically-observed program states where the bug template preconditions are met. Hence, the guarantees provided by \bi{} are relative to the correctness of the bug template specification. 
% Consequently, the bug templates must be created with care.
%That is, can the bugs in the provided list manifest in at least one execution of the program? LAVA~\cite{dolan2016lava} provides backtraces for each test case showing that the bugs included can actually be executed. EvilCoder \cite{pewny2016evilcoder}, however, provides no such guarantees. \bi{} benchmarks come with inputs which can generate dynamically-observed program states where the required preconditions (and postconditions when provided) are met. Thus, the guarantees provided by \bi{} are relative to the correctness of the bug template specification. Consequently, it is important for the user to create bug templates with care.

\paragraph{Matching ground truth to tool output} \label{subsec:matching}
Ground truth must include information such as location and bug type for each listed bug. This information allows automated or semi-automated matching of a tool's output with the ground truth. There are various pitfalls in providing this information: there may be multiple locations associated with a bug, multiple bug types associated with the same bug, multiple bugs in the same location (depending on the granularity of the location), or lexically distinct languages used by tools to warn about the same bug type. Several recent studies~\cite{Herter2017,delaitre15evaluating} elaborate on these problems.

\paragraph{Real-world bug distribution} \bi{} gives us  control over how many of each type of bug we inject. By injecting bugs of a type that are harder or easier for a given tool to detect, one can influence the measured recall of the tool on the generated benchmark. Unfortunately, it is difficult to know the real-world distribution of different bug types. This does not prevent the use of \bi{} for comparing the relative recall of two tools on particular bug types of interest or between different settings of the same tool.

%The Juliet test cases \cite{nistsard} come with a line number and a specification of the CWE that the bug belongs to.

LAVA~\cite{dolan2016lava} injects only buffer overflows, so the bug type is known up front. Every test case includes a backtrace that showcases the bug. While this may be sufficient for evaluating fuzzers or manually inspecting static analysis results, it can be difficult to automate. For example, do you credit a tool with finding a bug only if it warns about the location at the top of the backtrace, or is it sufficient for it to warn about any location in the backtrace? Are there other relevant locations in the program that can be justifiably reported by static analysis tools? For the LAVA-1 dataset, we found empirically that key locations in the backtraces can be matched to invocations of the synthetic method \texttt{lava\_get()} in the source code. Consequently, we interpret the ground truth to be the set of these locations.

For our \bi{} benchmark, such additional ground truth information is implicit in the bug templates (which specify the bug type) and the locations where the injection was performed. As shown in the example in \autoref{fig:scion-injection}, the injection location can be determined by examining the source code difference between the original and injected program.

A further hurdle to automation is that there is no standardized format for the output of a static analysis tool that all tools adhere to, and often no direct way to determine which specific bug a tool is reporting.
In practice, the evaluator must typically rely on manually created heuristics that match the tool's reports with ground truth based on location and warning type. This approach has some limitations, notably the possibility of mistakenly failing to credit the tool with a true positive because it reports a slightly different but related bug, or because it reports the correct bug at a slightly different location. Adding some ``tolerances'' to the location heuristics, such as allowing a neighborhood of several lines of code around the expected bug location, can mitigate this problem but may cause its own issues if the tool detects unrelated bugs in the neighborhood. In our experimental evaluation~\autoref{subsec:experimental-setup}, we explicitly discuss how we credit tools for finding appropriate bugs in our benchmarks.

%% file: texfiles/test-suite.tex
\section{Our benchmark suites}
\label{sec:testsuite}

In this section, we describe two \bi{}-generated benchmark suites. 
Both these benchmark suites, and the bug templates used to generate them, are available online for use by the community~\cite{buginjectortestsuite}. 
We plan to maintain a library of bug templates that can be used for different user-chosen evaluations. 

\subsection{Selection of host programs}

\begin{table}[h]
  \caption{\label{host-programs}Host programs used for evaluation. LOC gives the lines of code in the programs. The rest of the columns are described in ~\autoref{subsec:perf-characteristics}.}  
  \begin{center}
    \begin{tabular}{l r >{\raggedleft\arraybackslash}p{0.9cm} >{\raggedleft\arraybackslash}p{0.9cm}  >{\raggedleft\arraybackslash}p{0.9cm} >{\raggedleft\arraybackslash}p{0.9cm}}
      \toprule
      \textbf{Project} & \textbf{Version} & \textbf{LOC} & \textbf{Prep Time} & \textbf{Query Time} & \textbf{Sites/ KLOC} \\
      \midrule
      grep \cite{grep}         & 2.0    &  12K &   66 & 1.76  & 372.76 \\
      % lighttpd \cite{lighttpd} & 1.4.49 &  72,904 &  580 &  36.44 &  4.03 \\
      nginx \cite{nginx}       & 1.13.0 & 178K &  766 &  5.03 & 7.62 \\
      % sqlite \cite{sqlite}     & 3.21.0 & 218,401 & 1322 &  90.13 & 20.91 \\
      \bottomrule
    \end{tabular}
  \end{center}
  \iftight
  \vspace*{-2em}
  \fi
\end{table}

We use the open-source projects listed in \autoref{host-programs} as the host programs for generating our benchmark suites. 
We have successfully injected bugs into other C/C++ host programs (total of $15$ real-world programs to date), but we have not included them in this paper due to resource constraints in running experiments (\autoref{sec:evaluation}). 
One such excluded program is WireShark version 1.12.9, which has 2.3 million lines of code: it is the largest program we have successfully injected bugs into.
This demonstrates \bi{}'s ability to inject into a variety of real-world projects. 
% While \texttt{sqlite} was the largest program utilized here, we have had success on other programs in excess of 1 million lines of code (LOC). Because our bug templates made use of integer and string variables, host programs which contained a large number of these variables, such as \texttt{grep} and \texttt{sqlite}, were particularly well-suited for injection. Others, such as \texttt{nginx}, which lack a large number of integer variables, yielded fewer overall injections.
An important criteria for picking host programs is the availability of test suites with good code coverage: they provide a large number of distinct trace points for \bi{}, improving the chances of finding many suitable injection points by matching preconditions.

\subsection{Selection of bug templates}
\label{subsec:scionssource}

\begin{table}[h]
  \begin{center}
   \begin{tabular}{ l r r r r }
     \toprule
     \multirow{2}{*}{\thead{Bug Template\\Source}} & \multirow{2}{*}{\thead{No. of\\Templates}} & \multicolumn{3}{c}{mean counts} \\
     \cmidrule{3-5}
      & & \thead{LOC} & \thead{FVars} & \thead{CF~Stmts}\\
     \midrule
      CSA~\cite{csacheckers}         & 10 &  3.1 & 0.8 & 0.2 \\ % 1.23
      % CSA regression tests~\cite{csacheckerstests} & 55 &  3.56 & 0.91 & 0.40 \\ % 1.44
      Infer~\cite{fbinferbo,fbinfertests}         & 6  &  4.2 & 0.8 & 0.8 \\ % 0.00
      Juliet tests~\cite{juliettestsuite}          & 55 &  7.8 & 1.3 & 0.9 \\ % 1.11
      \bottomrule
    \end{tabular}
  \caption{\label{scion-statistics}The number of bug templates from each source. The last three columns provide the means over each set of templates for: (a) the number of lines of code to be injected, (b) the number of free variables to be rebound, and  (c) the number of control-flow statements in the injected code, respectively.}
  \end{center}
  \iftight
  \vspace*{-2em}
  \fi
\end{table}

We create bug templates from three sources (shown in \autoref{scion-statistics}) to satisfy two different goals. 
First, we want our benchmark suite to allow a fair evaluation of CSA~\cite{csa} and Infer~\cite{fbinfer}, and inject bug types that these tools care about and are expected to find. 
Both tools support the detection of buffer overflows (BO) and null pointer dereferences (NPD). 
Therefore, we collect examples of BO and NPD bugs that appear in these tool's documentation~\cite{csacheckers,fbinferbo} and regression test suites \cite{fbinfertests}. 
We manually verified that each example contains the bug they claim to contain, and then converted the example to a bug template. 
We also checked that at least one tool warns on each example bug snippet. 
The manual conversion of a bug example to a bug template is fairly straightforward (described in \autoref{bug-templates}), and only took on the order of few minutes per example.
Each of the $16$ bug templates collected from CSA and Infer are injected upto $30$ times into each of the two host programs in~\autoref{host-programs}, to create benchmark suite \textbf{B1}, with a total of $591$ program variants. 
Note that a bug template may have been injected fewer than $30$ times into a host program because of insufficient number of precondition-matching locations or failed validation.
In \textbf{B1}, each of the $16$ bug templates has been injected at least once. \textbf{B1} is used for answering the research questions (\autoref{subsec:research-questions}) \textbf{RQ1}, \textbf{RQ2}, and \textbf{RQ3}. 

Second, we want to demonstrate that \bi{} can inject a wide variety of bug types and CWE categories~\cite{cwe}. 
To this end, we \emph{automatically} converted $55$ bug examples from the Juliet test suite (version 1.3.~\cite{juliettestsuite}) into bug templates; these bug examples span 55 unique CWE types, from stack-based buffer overflows (CWE-121) to type confusion (CWE-843). 
We exploited the uniform structure of Juliet tests to automatically create these bug templates: we extract free variables, preconditions, and code to inject from the Juliet test suite using both static and dynamic information from each bug example.
We created the benchmark suite \textbf{B2}, which contains 2,492 program variants, by injecting each of the $55$ bug templates sourced from Juliet tests upto $30$ times into each of the host programs. 
Each of the 55 bug templates has been injected at least once. 
Many bug types in Juliet tests are out of scope for CSA and Infer, therefore we do not evaluate these tools on \textbf{B2} in this paper.
\textbf{B2} serves to answer \textbf{RQ4}.

The program variants with bugs are uniformly formatted using a code beautification tool, ensuring the injection does not stand out due to code-style differences. 
As shown in \autoref{scion-statistics}, the bug templates typically include a small amount of code. 
These characteristics, along with the use of existing program variables (through free variable rebinding), allow the injections to meld with the existing code and look realistic (e.g., see~\autoref{diff}, or examine any of the benchmark programs).

\subsection{Performance of \bi{}}
\label{subsec:perf-characteristics}

As discussed in~\autoref{sec:bi-arch}, \bi{} operates in a pipeline of several stages. Performance in these stages depends on characteristics of the host program, its tests, and the bug template set. \autoref{host-programs} summarizes the key characteristics and performance data for the host programs. 
Timing experiments were performed on an Intel(R) Xeon(R) 2.10 GHz machine with 72 cores and 128 GB of RAM. 

In the {\em instrument} and {\em execute} stages, \bi{} parses the host program, adds instrumentation, and runs the program with test inputs to collect traces.
The time required for this stage depends on the size of the program, the number of variables it contains, and the number of input tests to run; the ``Prep Time'' column in \autoref{host-programs}, given in seconds, provides this information for each host program. 
This provided prep time is a one time cost, which gets amortized over the number of bugs to be injected into the same host program.

The {\em inject} stage involves searching the trace database for points satisfying the bug template preconditions. 
The time required per injection depends on the number of points collected in the trace, the percentage of points which satisfy the precondition and free variable requirements, as well as the complexity of the precondition.
The ``Query Time'' column gives the median time (in seconds) per query.
The ``Sites/KLOC'' column in \autoref{host-programs} provides the number of matching host-program sites that are suited for injection based on our bug templates, per $1000$ lines of code. 
The \texttt{grep} program contained a large number of string and integer variables, and therefore showed higher density of potential injection sites; conversely, \texttt{nginx}, with few integer variables, had lower density of injection sites.

Lastly, \bi{} edits the program, applies code formatting to the buggy software, and writes it out to disk. The time required to apply code formatting and printing the buggy program is directly proportional to the program size. 
Overall, the prep time dominates the pipeline as the most expensive stage. 
Given the offline and automatic nature of benchmark creation, we believe the performance of \bi{} is reasonable. 
% Compared to the current process of manual benchmark creation, 

%% file: texfiles/evaluation.tex
\section{Evaluation}
\label{sec:evaluation}

In this section, we outline the research questions that direct our evaluation, describe our experimental methodology, report and discuss the results of our experiments, and compare our benchmark with the LAVA test cases~\cite{lavacorpora}.

\subsection{Research questions}
\label{subsec:research-questions}

The goal of our evaluation is to answer the following research questions about \bi{} and its generated benchmarks.
\begin{description}%[label=RQ\arabic*.]
  \item[RQ1:] \emph{Do the benchmarks contain bugs which are seemingly in scope for the tool but which the tool fails to detect?} Such bugs could provide useful concrete feedback to the tool's developers.
  \item[RQ2:] \emph{Can the benchmarks discriminate between different static analysis tools?} %Such a discrimination allows for showcasing each tool's strengths and weaknesses.
        % \item can discriminate between different versions of a static analysis tool, that is, demonstrates situations where an earlier version of a tool fails to find a bug but a later, improved version does find it. This would be an indication that the suite is useful for identifying specific weaknesses in a tool that developers can target for improvement.
  \item[RQ3:] \emph{Can the benchmarks discriminate between different parameter settings for a given static analysis tool?} Such an ability suggests the use of \bi{} for automated tuning of a tool's parameters specific to a given codebase.
  \item[RQ4:] \emph{Can \bi{} create benchmarks that include bugs from multiple CWEs?} Such an ability shows whether the technique is applicable to multiple bug types.
\end{description}

In addition to answering the above research questions, we also compare \bi{} with the LAVA test suite with respect to the same research themes.

\subsection{Experimental setup and methodology}
\label{subsec:experimental-setup}

\paragraph{Static analysis tools and configurations} We perform our experiments using two open-source state-of-the-art static analysis tools for C/C++ programs: Clang Static Analyzer (CSA)~\cite{csa} and Infer~\cite{fbinfer}.
We use CSA version 3.8\footnote{This is the default version available on Ubuntu 16.04.}, and run the tool on Ubuntu 16.04.
We use CSA with the analyzer configuration mode set to ``shallow'' (CSA-S), as well as the default mode ``deep'' (CSA-D). 
CSA-S mode changes certain default analysis parameters, such as the style of the inter-procedural analysis and maximum inlinable size.
% Thus, CSA-S 9.0 refers to version 9.0 of CSA run in shallow mode. 
We use the term CSA to refer to both modes. 
CSA is run with all the default checkers enabled, along with the optional \texttt{alpha}, \texttt{security}, \texttt{osx}, \texttt{llvm}, \texttt{nullability}, and \texttt{optin} checkers.

We use Infer version 0.14, and run it from the tool's official Docker image. 
Infer is run with default options and \texttt{compute-analytics}, \texttt{biabduction}, \texttt{quandary}, and \texttt{bufferoverrun} enabled.
For both CSA and Infer, our intention is to enable as many checkers as possible to maximize the tool's chance of finding the injected bugs.

\paragraph{Projected recall} This metric computes the percentage of the \bi{} injected bugs found by a tool.
We report this value by rounding to a whole number percentage.
To determine whether a tool found an injected bug successfully, as discussed in \autoref{subsec:matching}, we consider the locations of the bug injection as the bug locations. We credit a tool with finding an injected bug if it reports a bug of an appropriate type on at least one of the injected code lines. The table below summarizes which tool-specific bug types (cell contents) reported by the tools (row headers) are considered to correspond to the injected bug types (column headers). We interpret the bug types reported by the tools quite generously, to maximize their chances of being credited with finding the injected bugs.

\begin{table}[h]
  \centering
  \begin{tabular}{ p{0.3cm}  p{3.3cm}  p{3.8cm} }
    \toprule
                   & \textbf{Buffer overrun (BO)}                                                                            & \textbf{Null pointer dereference (NPD)}                                                                 \\
    \midrule
    \textbf{CSA}   & Out of bound array access, Result of operation is garbage or undefined, \texttt{malloc()} size overflow & Dereference of null pointer, Uninitialized argument value, Argument with `nonull' attribute passed null \\
    \midrule
    \textbf{Infer} & Array out of bounds, Buffer overrun, Memory leak, Stack variable address escape                         & Array out of bounds, Buffer overrun, Dangling pointer dereference, Null dereference, Memory leak        \\
    \bottomrule
  \end{tabular}
  % \caption{\label{tab:tool-bug-type} Summary of bug types considered for each evaluated tool. These are all the bug types for a given tool (row headers) that correspond to the types (column headers) of injected bugs. }
  % \iftight
  % \vspace*{-1em}
  % \fi
\end{table}

\subsection{Experiments and results}
\label{subsec:exp-and-results}

To help answer research questions \textbf{RQ1}, \textbf{RQ2}, and \textbf{RQ3}, we run CSA-S, CSA-D, and Infer, on benchmark \textbf{B1} (described in \autoref{sec:testsuite}).
Tables~\ref{scion-source-summary} and~\ref{scion-kind-summary} provide  the projected recall of the tools on various partitionings of \textbf{B1}.

The two tables provide different views of the same experimental results. 
\autoref{scion-source-summary} partitions the results by bug template source (\autoref{scion-statistics}) and host program (\autoref{host-programs}). 
The last row provides results on the entire \textbf{B1} benchmark.
\autoref{scion-kind-summary} partitions the results by injected bug type (NPD or BO) and host program.
The ``No. of Bugs'' column in both these tables describes the number of  benchmark programs--each containing one known bug---in the specified partition.
The rightmost six rows in both the tables provide the projected recall of the tools on the given benchmark partition.
The highest projected recall in each row is highlighted.

\begin{table}
\caption{\label{scion-source-summary} Projected recall of different tools for various pairs of bug template sources and host programs.
Last row summarizes the results over the entire benchmark suite \textbf{B1}.}   
\centering
\input{texfiles/scion-source-summary-table.tex}
\end{table}

\begin{table}
\caption{\label{scion-kind-summary} Projected recall of different tools for various pairs of bug types and host programs.} 
\centering
\input{texfiles/scion-kind-summary-table.tex}
\end{table}

A tool can obtain higher projected recall by simply reporting more warnings overall, which will increase its chances of also reporting an injected bug.
A tool could also take a lot longer than is acceptable to a user to report bugs. 
Therefore, it is instructive to look at two additional metrics: ``Warnings per KLOC'' and ``Time taken''. \autoref{warnings-per-kloc} reports these metrics. 

\begin{table}
\caption{\label{warnings-per-kloc} Total warnings reported per KLOC and average time taken by the tools.} 
  \begin{center}
    \begin{tabular}{l r r r r r r}
      \toprule
      \multirow{2}{*}{\thead{Host}} & \multicolumn{3}{c}{Warnings per KLOC} & \multicolumn{3}{c}{Time taken (seconds)} \\
      & \thead{CSA-S} & \thead{CSA-D} & \thead{Infer} & \thead{CSA-S} & \thead{CSA-D} & \thead{Infer} \\
      \midrule
      grep & 7.9$\pm$.2 & 9.7$\pm$.2 & 1.9$\pm$.1 & 14.7 & 41.1 & 20.5   \\
      nginx & 6.8$\pm$.0 & 2.8$\pm$.0 & 0.8$\pm$.1 & 229.6 & 366.1 & 338.7 \\
      \bottomrule
    \end{tabular}
  \end{center}
\end{table} 

The columns under ``Warnings per KLOC'' in \autoref{warnings-per-kloc} provide the average number of total warnings reported by the tools for every thousand lines of code, on the benchmark suite \textbf{B1}. 
The number provided after the $\pm$ symbol is the standard deviation (rounded to one decimal place) over all variants of that host program.
This metric can be helpful to check that a tool is not reporting so many warnings on real-world programs that it is effectively unusable.  Note that comparing this metric directly between two analysis tools which do not have comparable warning classes (such as Infer vs. CSA) is not particularly meaningful.
On ``grep'', CSA-D reports more total warnings than CSA-S, whereas, on ``nginx'', CSA-S reports more total warnings than CSA-D. 

The columns under ``Time taken'' in \autoref{warnings-per-kloc} provide the average time taken by the tool to run on a given host program. We run each tool five times on a four-core Intel(R) Xeon(R) 2.10Ghz machine with 16GB RAM and report the average. CSA-S runs much faster than CSA-D. 

%% per scion data
% count	mean	std	min	25%	50%	75%	max
% scion								
% CLANG-BUFFER1	6.0	0.644444	0.499185	0.000000	0.241667	0.966667	0.966667	0.966667
% CLANG-BUFFER1-TRIVIAL	11.0	0.560606	0.450634	0.000000	0.000000	0.833333	0.916667	1.000000
% CLANG-BUFFER2	11.0	0.481818	0.479204	0.000000	0.000000	0.600000	1.000000	1.000000
% CLANG-BUFFER2-TRIVIAL	11.0	0.860606	0.097546	0.700000	0.800000	0.866667	0.933333	1.000000
% CLANG-BUFFER3	6.0	0.850000	0.164317	0.600000	0.750000	0.900000	0.975000	1.000000
% CLANG-BUFFER3-TRIVIAL	11.0	0.860606	0.097546	0.700000	0.800000	0.866667	0.933333	1.000000
% CLANG-BUFFER4	6.0	0.666667	0.516398	0.000000	0.250000	1.000000	1.000000	1.000000
% CLANG-POINTER-DEREFERENCE1	6.0	0.605263	0.477470	0.000000	0.184211	0.842105	0.947368	1.000000
% CLANG-POINTER-DEREFERENCE2	6.0	0.500000	0.456107	0.000000	0.078947	0.526316	0.894737	1.000000
% CLANG-POINTER-DEREFERENCE3-TRIVIAL	11.0	0.739394	0.227459	0.400000	0.533333	0.833333	0.933333	0.966667
% INFER-BUFFER1	11.0	0.363636	0.504525	0.000000	0.000000	0.000000	1.000000	1.000000
% INFER-BUFFER2	11.0	0.363636	0.504525	0.000000	0.000000	0.000000	1.000000	1.000000
% INFER-REGRESSION-TEST-BUFFER1	11.0	0.721212	0.253978	0.333333	0.533333	0.866667	0.900000	1.000000
% INFER-REGRESSION-TEST-BUFFER2	6.0	0.111111	0.173419	0.000000	0.000000	0.000000	0.225000	0.366667
% INFER-REGRESSION-TEST-BUFFER3	11.0	0.304075	0.383308	0.000000	0.000000	0.103448	0.758621	0.827586
% INFER-REGRESSION-TEST-POINTER-DEREFERENCE1	11.0	0.572727	0.457971	0.000000	0.000000	0.833333	0.933333	1.000000

\paragraph{Addressing RQ1} 
The benchmark suite \textbf{B1} consists of the bugs that CSA and Infer care about injected into popular open-source programs (representing the expected targets of the chosen static analysis tools). 
Tables~\ref{scion-source-summary}~and\ref{scion-kind-summary} show that both the tools detect some but not all of the injected bugs. 

If a tool reports a bug on a small example with a simple context, we might expect that the tool also reports a similar bug in a more complex setting. 
However, in the case of both CSA and Infer---the leading open-source static analysis tools for C/C++---we find that they ``lose'' bugs (projected recall is not $100\%$) across all rows in Tables~\ref{scion-source-summary}~and\ref{scion-kind-summary}. 
That is, CSA and Infer find bugs in their respective documentation examples and regression tests, but in many cases they lose the ability to find the ``same'' bug when it is injected and integrated into a larger program. 
These ``lost'' bugs can represent concrete feedback for the analysis tool developers.

\paragraph{Addressing RQ2} 
Tables~\ref{scion-source-summary}~and\ref{scion-kind-summary} show that our generated benchmarks can be used for contrasting the projected recall of the evaluated tools. 
Depending on the specific subset of the benchmark suite (i.e., table row) that is of interest to the evaluator, different tools have higher projected recall. 
Thus, \bi{} can be used to perform tool evaluations to suit specific customer needs by providing control over the distribution of bug templates and host programs. 
CSA-S has the highest recall on benchmark suite \textbf{B1} as a whole. 

\paragraph{Addressing RQ3} 
Static analysis tools are typically configurable, with the chosen configuration affecting tool recall, precision, and scalability. There is generally no single best configuration: it depends on several factors including the codebase being analyzed, the warning classes that are of interest, etc. 
To evaluate how our generated benchmarks discriminate different configurations of the same tool, we examine two configurations of Clang Static Analyzer: CSA-D and CSA-S. 
In a majority of cases in \textbf{B1}, CSA-S has equal or higher projected recall compared to CSA-D, while also being significantly faster. 
% Thus, on this particular benchmark, running CSA with \texttt{mode=shallow} instead of the default setting seems to be a better choice. 
This is a surprising result that may be of interest to CSA users and developers.

In this paper, we only compare two configuration points of CSA. 
However, CSA (and many other analysis tools) have several configuration parameters. 
Projected recall from \bi{} generated tests can be used (in conjuction with other metrics of interest) to tune the settings of these parameters for a given codebase.

\paragraph{Addressing RQ4} 
\bi{} is able to generate the benchmark \textbf{B2}, which contains injected bugs corresponding to $55$ different CWEs, based on bug templates sourced from Juliet test suite version 1.3. 
This artifact shows that \bi{} can be used to inject a wide-variety of bug types.

\subsection{Causes of lost bugs}
\label{subsec:lost-bugs}

A large number of injected bugs are ``lost'' by the evaluated tools (ranging between $28\%$ to $54\%$ lost bugs per tool). 
An extensive study of all lost bug cases by each tool is out of scope for this work.
Instead, we sampled a small number of randomly-selected lost bugs to manually check whether there were particular patterns or language constructs that were causing the tools to lose track of the bugs.
However, we found no single dominant pattern for the lost bugs: there seems to be a long tail of several issues that cause tools to lose bugs.
In our limited study, we see that each lost bug belongs to one of three categories: 
\begin{description}
  \item [needs-fix:] the tool needs to be fixed for the bug to be found
  \item [param:] adjusting the tool's parameters can find the bug
  \item [expected:] the bug is lost by design
\end{description}

Below, we discuss some simplified examples of lost bugs in \textbf{B1}.
We mark each discussed bug with our diagnosis with respect to the above categories.
We have reported~\cite{InferBugReport1,InferBugReport2,CSABugReport1} some of these lost bugs to the analysis developers.

Infer fails to report~\cite{InferBugReport2} any warnings in a function that has enum declarations of the following form (\textbf{needs-fix}):
\begin{lstlisting}[language=C]
  enum { L, R } dirs[12];
\end{lstlisting}
Infer fails to report~\cite{InferBugReport1} the null pointer dereference in this simple case (\textbf{needs-fix} or \textbf{param}):
\begin{lstlisting}[language=C]
  int *nullable; int *firstpos; int *lastpos;
  int * buf = 0;
  nullable = malloc(2*sizeof(int));
  firstpos = malloc(2*sizeof(int));
  lastpos = malloc(2*sizeof(int));
  for (int i = 0; i < 3; i++) { /* no-op */ }
  *buf = 1; // null pointer dereference
  nullable++; firstpos++; lastpos++;
\end{lstlisting}
Infer fails to report any bugs present in the source code of those functions for which library models exist (\textbf{expected}).
The source code of such functions are ignored. 
This behavior may result in supply-chain attacks going unnoticed by the tool. 

CSA fails to report~\cite{CSABugReport1} a buffer overrun in the presence of an intervening function call, presumably due to unsound early termination in the tool's path exploration (\textbf{needs-fix}). 

Many of the bugs lost by CSA can be found by tuning the analysis parameters (\textbf{param}). 
E.g., high values of the parameter \texttt{-maxloop}, which controls the number of times a block can be visited before giving up, finds many lost bugs.
% CSA version 9.0 with the the ``deep'' mode crashes (\textbf{needs-fix}) when running on injections into the ``nginx'' host program.
% None of the other CSA variants crash.

Thus, \bi{} generated benchmarks can expose the various real-world scenarios in which an analysis tool can fail to report a bug, which is of interest to both analysis users and analysis developers.

% We found that without a more specific understanding of analysis implementation internals, we cannot pin-point a set of reasons why the analysis tools lost certain bugs.
% In some of the bugs lost by CSA, the buggy behavior involved local reasoning, but the buggy code was inside loops with no explicit terminating condition, with loop exit only through conditional goto or break statements. Some of the other lost bugs involved data/control dependence on global variables. But none of these are consistent patterns, e.g., CSA does find bugs when global variables are involved in many cases.

% \begin{figure}
% \begin{lstlisting}[language=C]
% static void prtext(/*some params*/ int *nlinesp) {
%   /* elided host program code */
%   if (!out_quiet) {
%     bp = lastout ? lastout : bufbeg;
%     /* from input (./harness.sh test BIN 5) */
%     if (!nlinesp) {
%       /* POTENTIAL FLAW */
%       *nlinesp = 0;
%   /* elided host program code */
% \end{lstlisting}
% \caption{\label{fig:lost-bug-code-example} Example code lost by CSA, detected by CSA-S.}
% \iftight
% \vspace*{-1em}
% \fi
% \end{figure}

% \autoref{fig:lost-bug-code-example} shows an injected bug that is detected by CSA-S but lost by CSA. The bug is the null pointer dereference on the line after the comment ``POTENTIAL FLAW''. It is a real bug, as it can be triggered with the provided input test.

\subsection{Comparison with LAVA benchmarks}
\label{subsec:lava}

We run CSA and Infer tools on the LAVA-1 benchmarks. The LAVA-1 benchmarks consist of $69$ variations of the \texttt{file} program, with each variant having one injected buffer overflow bug. As discussed in \autoref{sec:foundations}, we stipulate for the sake of this evaluation that the bug location is the line consisting of a \texttt{lava\_get()} call, and give a tool credit for identifying the bug if it specifies a location within 5 lines of this. In all LAVA-1 test cases, we found that the \texttt{lava\_get()} call location matched the first location provided in the corresponding backtrace included with the LAVA corpus.

CSA reports between $41$ and $51$ warnings on each of the LAVA-1 benchmarks. In $58$ of the $69$ programs, CSA does not report on any LAVA-injected bugs. In the remaining $11$ programs, CSA issues warnings at the injected bug locations. Upon manual inspection of each of these examples, we determined these warnings to be unrelated to buffer overflows.\footnote{The reported warnings were one of: ``pointer of type \texttt{void*} used in arithmetic'', ``nested extern declaration of \texttt{vasprintf}'', ``implicit declaration of function \texttt{vasprintf}'',  ``pointer arithmetic on non-array variables relies on memory layout: which is dangerous''.} 

Infer reports between $16$ and $18$ warnings on each of the LAVA-1 benchmarks. However, none of the Infer warnings are at the LAVA-injected bug locations.

\begin{figure}
\begin{lstlisting}[language=C]
/* inside a for loop */
if (ml->map)
  apprentice_unmap(((ml->map))+(lava_get())*((0x12345678
    <= (lava_get()) && 0x123456f8 >= (lava_get())) ||
    (0x12345678 <= __bswap_32((lava_get())) && 0x123456f8
    >= __bswap_32((lava_get())))));
free(ml);
\end{lstlisting}
\caption{\label{fig:example-lava-snippet}Example LAVA-injected bug.}
\iftight
  \vspace*{-1em}
\fi
\end{figure}

To summarize, both CSA and Infer report warnings on the LAVA-1 benchmarks, but none of these are related to the LAVA-injected bugs. Thus, the projected recall of both of these tools is \textbf{$0\%$} on the LAVA-1 benchmarks. This result is not particularly surprising, as LAVA is biased towards testing the limits of fuzzing tools, and injects code that looks like the snippet in~\autoref{fig:example-lava-snippet}. Such bugs would typically be out of scope for accurate reasoning by most static analysis tools, as the tools have to make static approximations and/or heuristic choices that balance precision, recall, and scalability. These results---that leading open-source static analysis tools have \textbf{zero} projected recall---indicate that LAVA benchmarks are not well-suited for discriminating between different static analysis tools (refer \textbf{RQ2}), or that they include bugs that are in scope for the evaluated static analysis tools (refer \textbf{RQ1}). Also, LAVA can only inject a very small number of bug kinds (refer \textbf{RQ4}).

Therefore, while LAVA has been successful in advancing fuzzing techniques~\cite{Rawat2017} and helping create capture-the-flag-style competitions~\cite{Hulin2017}, it is less relevant in evaluating static analysis tools.

%% file: texfiles/scion-source-summary-table.tex
\begin{tabular}{c c r r r r}
  \toprule
  \thead{Bug\\Template\\Source} & \thead{Host\\Program} & \thead{No. of\\Bugs} & \thead{CSA-S} & \thead{CSA-D} & \thead{Infer} \\
  \midrule
  \multirow{2}{*}{CSA} & grep & 251 & \hilightcell{88\%} & 69\% & 45\% \\
   & nginx & 122 & \hilightcell{92\%} & \hilightcell{92\%} & 52\% \\
  \multirow{2}{*}{Infer} & grep & 179 & 36\% & 37\% & \hilightcell{50\%} \\
   & nginx & 39 & \hilightcell{69\%} & \hilightcell{69\%} & 18\% \\
  \midrule 
  Both & Both & 591 & \hilightcell{72\%} & 64\% & 46\% \\
  \bottomrule
\end{tabular}

%% file: texfiles/scion-kind-summary-table.tex
\begin{tabular}{c c r r r r}
  \toprule
  \thead{Bug\\Type} & \thead{Host\\Program} & \thead{No. of\\Bugs} & \thead{CSA-S} & \thead{CSA-D} & \thead{Infer} \\
  \midrule
  \multirow{2}{*}{NPD} & grep & 98 & \hilightcell{82\%} & 67\% & 15\% \\
    & nginx & 60 & \hilightcell{90\%} & \hilightcell{90\%} & 20\% \\
  \multirow{2}{*}{BO} & grep & 332 & \hilightcell{61\%} & 53\% & 57\% \\
    & nginx & 101 & \hilightcell{84\%} & \hilightcell{84\%} & 58\% \\
  \bottomrule
\end{tabular}

%% file: texfiles/threats-validity.tex
\section{Limitations and future work}
\label{sec:threats}

\bi{} currently chooses an injection point in the host program uniformly at random from all the dynamic trace points that match the bug template's preconditions. Thus, host program points that are exercised more frequently by the accompanying tests are more likely to be used for injection, as they appear more frequently in the dynamic traces. \bi{} can be combined with coverage-increasing input-generation techniques like concolic testing~\cite{Majumdar2007} to obtain an improved program-wide distribution of injected bugs. 

\bi{} does not currently support the injection of concurrency-related bugs. We plan to add such support. Our first step will be to improve instrumentation so that concurrency-related information such as the current thread and process is available in the trace.

\bi{} cannot always inject a bug template into a host program, because there is not always a dynamic trace point that matches all the preconditions and free-variable requirements for the template. To increase the chances of finding injection points in a host program, we plan to enhance \bi{} to allow for variable rebinding to aggregate structs and fields.

We envision running \bi{}'s pipeline multiple times in an evolutionarily-guided heuristic search. This process would allow injection of multiple bugs into a single host program, maximizing an objective function that balances factors such as number of injected bugs, naturalness of code~\cite{ray16naturalness}, realistic distribution of bugs~\cite{ostrand2002distribution,fenton2000quantitative}, retention of the original program behavior, and syntactic/stylistic similarity~\cite{caliskan2015anonymizing} between the buggy program and the original program. \bi{} is built using SEL, which supports evolutionary search with multi-objective fitness functions. Leveraging this support, we have early prototypes that fulfill this vision.

% Regarding our test suite described in \autoref{sec:testsuite}, the
% realism of the bug templates used in the paper should be on a par with
% the sources from which they were extracted. The main remaining challenge is providing a stronger guarantee of realism through a comparison with real-world software. As we discuss in \autoref{sec:relwork}, there is a large body of research on bug patterns and characteristics in real-world code. In future work, we plan to apply the patterns and metrics from this research to
% assess our test suite and to tune it towards greater realism. We note
% that the evolutionary-search enhancement of \bi described above can
% use quantitative information about bug patterns and/or distribution as
% part of its fitness function. This would allow it to generate a more
% realistic test suite.

Regarding our experimental methodology, the main threat to validity relates to how we measure whether a tool finds a specific injected bug, both for \bi and LAVA test cases. As we explain in \autoref{sec:foundations} and \autoref{sec:evaluation}, we use simple heuristics to match the location contained in the tool's warning with the location of the known bug and determine that the correct bug has been identified if the bug types match and the locations are within a certain maximum distance. We could refine this heuristic by using more sophisticated matching techniques from related work on the issue of deduplicating and/or clustering tool warning reports \cite{fry13clustering,muske16survey}.

%% file: texfiles/conclusion.tex
\section{Conclusion}
\label{sec:conclusion}

In this paper, we introduce \bi{}, a system that automatically generates bug-containing benchmarks suitable for evaluating and testing software analysis tools. 
\bi{} operates by injecting bug templates into real-world programs, and is able to create custom benchmarks that are real-world-like, can draw from a wide variety of bug types, and come with bug-triggering inputs. 
Our experimental evaluation shows that \bi{} benchmarks are useful for several purposes: (a) showcasing bugs that are seemingly in scope for a tool to find but that the tool misses, (b) discriminating between and guiding the improvement of static analysis tools, and (c) tuning tool parameters for a specific codebase.
We also show that \bi{} can create bugs from multiple CWEs.

%% file: texfiles/acks.tex
\section*{Acknowledgments}

This material is based on research sponsored by the Defense Advanced Research Projects Agency (DARPA) under Contract No. D17PC00096 and the Department of Homeland Security (DHS) Science and Technology Directorate, Cyber Security Division (DHS S\&T/CSD) via contract number HHSP233201600062C. The views, opinions, findings, and conclusions or recommendations contained herein are those of the authors and should not be interpreted as necessarily representing the official views policies or endorsements, either expressed or implied, of DARPA or DHS.
We would like to thank Jeff Foster, Mikael Lindvall, Paul Black, Daniel Krupp, Amy Gale, John Regehr, and the SAMATE group at NIST for their feedback on our work. 